# Total Differential Errors in One-Port Network Analyzer Measurements with Application to Antenna Impedance

*Nikolitsa YANNOPOULOU, Petros ZIMOURTOPOULOS*

Antennas Research Group
Dept. of Electrical Engineering and Computer Engineering, Democritus University of Thrace, Xanthi, Greece

**Abstract.** *The objective was to study uncertainty in antenna input impedance resulting from full one-port Vector Network Analyzer (VNA) measurements. The VNA process equation in the reflection coefficient ρ of a load, its measurement m and three errors Es -determinable from three standard loads and their measurements- was considered. Differentials were selected to represent measurement inaccuracies and load uncertainties (Differential Errors). The differential operator was applied on the process equation and the total differential error dρ for any unknown load (Device Under Test DUT) was expressed in terms of dEs and dm, without any simplification. Consequently, the differential error of input impedance Z -or any other physical quantity differentiably dependent on ρ- is expressible. Furthermore, to express precisely a comparison relation between complex differential errors, the geometric Differential Error Region and its Differential Error Intervals were defined. Practical results are presented for an indoor UHF ground-plane antenna in contrast with a common 50 Ω DC resistor inside an aluminum box. These two built, unshielded and shielded, DUTs were tested against frequency under different system configurations and measurement considerations. Intermediate results for Es and dEs characterize the measurement system itself. A number of calculations and illustrations demonstrate the application of the method.*

## 1. Introduction

In full one-port measurements with a vector network analyzer (VNA) of real characteristic impedance $Z_0$, a device under test (DUT) with impedance Z has a complex reflection coefficient ρ defined by

$$\rho = (Z - Z_0)/(Z + Z_0) \qquad (1)$$

and related to its complex VNA measurement m by a bilinear transformation

$$\rho = (m - D)/[M(m - D) + R] \qquad (2)$$

in which all the quantities are implicitly dependent on the frequency.

The quantities D, M and R have been defined as system errors Es and a physical meaning has been given to them [1]. Accordingly, D is the directivity error $E_D$, M is the source match error $E_M$ and R is the frequency response error $E_R$. Although it is possible to define Es in terms of elementary circuit quantities, as it has been analytically proven by the authors for typical VNA system configurations that will be described in the following, this analysis is too extensive to be reproduced here. The resulting equivalent error model is shown as flow graph in Fig. 1.

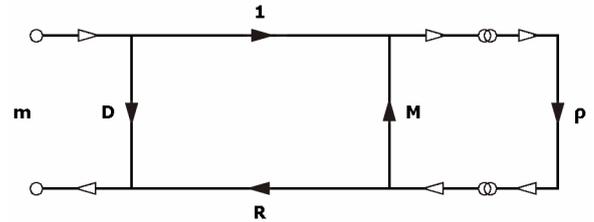

**Fig. 1.** Full one-port error model.

Mathematically, transformation (2) can be uniquely determined from three given distinct values $\rho = \rho_k$ (k = 1, 2, 3), with modulus $|\rho_k|$ and argument $\theta_k$, and respectively known $m = m_k$, with modulus $|m_k|$ and argument $\varphi_k$ [2]. This determination expresses D, M and R in terms of three standard reflection coefficient values and corresponding VNA measurements. Since ρ of any unknown DUT is calculated by (2), the measurement system itself is characterized mainly by the D, M and R in terms of frequency. Therefore, ρ is calculated from seven complex numbers: m, $m_k$ and $\rho_k$. However, since the four measurements have inaccuracy and the three standard loads uncertainty, there is an error (uncertainty) in the value of ρ. In addition, since ρ is a complex number, this error in ρ has a geometric representation as a region of the complex plane that may be used efficiently in comparison issues.

To the best of the authors' knowledge, for the ρ error estimation and its geometric representation:

(a) There are several numerical techniques based on the simplest approximation of ρ error by the $\Delta S_{11} \cong |m| - |\rho|$ equation of measurement uncertainty, which is graphically represented by a circle of radius $|\Delta S_{11}|$ around ρ [3]-[4].

(b) There are analytic methods using partial derivatives for specific or nonspecific $\rho_k$ values, to estimate the influence of one or more standard load uncertainties -but not of the inaccuracies of measurements- on the ρ error. The most complete of them is perhaps the work of Stumper, who studied full two-port VNA measurements in 2003 [5]. However, although full one-port measurements can be considered in general as a simplified application of two-port measurements, the partial deviations, for the three load uncertainties given in [5], cannot be generalized to include the four measurement inaccuracies.

(c) There is no analytic expression using the total differential dρ or method using its geometric representation.

In this paper, the complete expression of the total differential error dρ for the reflection coefficient ρ and its exact geometric representation are expressed, without any simplification due to a particular load value and/or a negligible load uncertainty and/or an insignificant measurement inaccuracy. Thus, the expression for the differential error of any physical quantity, differentiably dependent on the reflection coefficient, is made possible. This includes the case of the uncertainty of the input impedance Z that can always be expressed by (1) in terms of ρ and practically used, as long as a pair of input terminals can be well defined for the DUT.

## 2. Theory

The following form of (2) was considered as the process equation in five complex variables

$$\rho m M - \rho M D + \rho R - m + D = 0 \tag{3}$$

The application of the differential operator to (3) resulted in a process equation in five differentials

$$(1 - \rho M)dD + \rho(m - D)dM + \rho dR$$
$$+ [R + M(m - D)]d\rho + (\rho M - 1)dm = 0 \tag{4}$$

The equation (3) was applied three times for the three standard loads, with values of $\rho_k$ equal to A, B, C and their three VNA measurements $m_k$ equal to a, b, c, respectively. After that, the system of three process equations was solved for D, M and R

$$D = [abC(A - B) + bcA(B - C) + caB(C - A)]/F$$
$$= \sum abC(A - B)/F \tag{5}$$

$$M = [c(B - A) + a(C - B) + b(A - C)]/F$$
$$= \sum c(B - A)/F \tag{6}$$

$$R = [(A - B)(a - b)(B - C)(b - c)(C - A)(c - a)]/F^2$$
$$= [\prod (A - B)(a - b)]/F^2 \tag{7}$$

with $\quad F \equiv cC(B - A) + aA(C - B) + bB(A - C)$
$$= \sum cC(B - A)) \tag{8}$$

where $\sum$ and $\prod$ produce two more terms, from the given one, by cyclic rotation of the letters a, b, c or A, B, C. The determination of errors is known as the calibration of the VNA measurement system and the three standards A, B, C are called calibration standards.

These errors were considered as dependent on the variables a, b, c, A, B, C and thus the three process equations in differentials formed a system, which was then solved for the three differentials dD, dM and dR

$$dD = [\prod (a - b) \sum (B - C)BCdA +$$
$$+ \sum (b - c)^2(B - A)(C - A)BCda]/F^2$$
$$= (1/F^2)\{(a - b)(b - c)(c - a)$$
$$\cdot [(B - C)BCdA + (C - A)CAdB + (A - B)ABdC]$$
$$+ (b - c)^2(B - A)(C - A)BCda +$$
$$+ (c - a)^2(C - B)(A - B)CAdb +$$
$$+ (a - b)^2(A - C)(B - C)ABdc\} \tag{9}$$

$$dM = [\sum (a - b)(c - a)(B - C)^2 dA$$
$$- \prod (A - B) \sum (b - c)da]/F^2$$
$$= (1/F^2)\{(a - b)(c - a)(B - C)^2 dA$$
$$+ (b - c)(a - b)(C - A)^2 dB$$
$$+ (c - a)(b - c)(A - B)^2 dC$$
$$- (A - B)(B - C)(C - A)$$
$$\cdot [(b - c)da + (c - a)db + (a - b)dc]\} \tag{10}$$

$$dR = \{\sum [F + 2(a - b)B(A - C)][(B - C)^2 dA \prod(a - b)$$
$$- (b - c)^2 da \prod (A - B)]\}/F^3$$
$$= (1/F^3)\{[F + 2(a - b)B(A - C)]$$
$$\cdot [(a - b)(b - c)(c - a)(B - C)^2 dA$$
$$- (A - B)(B - C)(C - A)(b - c)^2 da]$$
$$+ [F + 2(b - c)C(B - A)]$$
$$\cdot [(a - b)(b - c)(c - a)(C - A)^2 dB$$
$$- (A - B)(B - C)(C - A)(c - a)^2 db]$$
$$+ [F + 2(c - a)A(C - B)]$$
$$\cdot [(a - b)(b - c)(c - a)(A - B)^2 dC$$
$$- (A - B)(B - C)(C - A)(a - b)^2 dc]\} \tag{11}$$

The developed expressions (9)-(11) are the total differential errors for the system errors D, M, and R. These expressions were mechanically verified using a developed software program for symbolic computations.

Notably, using manufacturer's data for standard load uncertainties and VNA measurement inaccuracies, the characterization of the measurement system can be completed by considering dD, dM and dR in terms of frequency and, perhaps, we can call the set of them "the differential error core of the measurement system".

The total differential error of $\rho$ was then expressed by

$$d\rho = [-RdD - (m - D)^2 dM - (m - D)dR + Rdm] / [M(m - D) + R]^2 \quad (12)$$

which was considered dependent, through dD, dM and dR, on L = 7 independent variables m, $m_k$ (a, b, c), $\rho_k$ (A, B, C) and on their L = 7 independent differentials dm, $dm_k$ (da, db, dc), $d\rho_k$ (dA, dB, dC).

To make possible a precise comparison between various complex differential errors, geometric notions were introduced below.

Since $Z_0$ is real, (1) is a transformation of the close right half plane to the closed unit circle [6]. Therefore, if $|\rho| = 1$, care must be exercised to restrict its differential into the unit circle. The VNA measurements have a specific bounded range for their modulus away from the origin O of the complex plane, so that the domain of each measurement is a bounded circular annular with its centre at the O.

Uncertainty and inaccuracy data outline regions for each d$\rho$ and its dm. If $z = |z|e^{jy}$ stands for any of the independent variables and $dz = e^{jy}(d|z| + j|z|dy)$ for its differential (where d|z| and dy in dz polar form are independent real differentials with values in given intervals) then the corresponding contribution to d$\rho$ is a summation term Wdz, with factor $W = |W|e^{jV}$, so that

$$Wdz = |W|e^{j(V + y)}d|z| + |W|e^{j(V + y + \pi/2)}|z|dy \quad (13)$$

W is in fact a known value of the respective partial derivative. Each expression Wdz outlines a contour for a partial z Differential Error Region (z DER) around O. If $z \neq 0$, the partial DER is a parallelogram with perpendicular sides d|z| and |z|dy, initially parallel to the rectangular coordinate axes Re{z} and Im{z}, stretched or contracted by |W| and rotated by (V + y) around O. If $z = \rho = 0$ then $dz = e^{jy}d|z|$, with $0 \leq d|z|$ and indeterminate y, so that the corresponding partial DER is a circle with radius |W|d|z|.

Accordingly, a total DER is the sum of either L parallelograms or (L − 1) parallelograms and 1 circle. DER is then a convex set with contour: either a polygonal line of 4L line segments and vertices, at most or a piecewise curve composed of 4(L − 1) line segments, 4(L − 1) circular arcs and 8(L − 1) vertices, at most. Some vertices may coincide.

Differential Error Intervals DEIs were defined by the greatest lower and least upper differential error bounds for the real and imaginary parts of d$\rho$. DEIs are the projections of a DER on the rectangular coordinate axes. In other words, DEIs are the sides of the approximate upright rectangle which is circumscribed to a given exact DER.

On the occasion: On the one hand, the commonly used approximation, mentioned in 1(a), is related to the exact maximum modulus of the differential error |d$\rho$|, which in fact is the radius of the circle that circumscribes the $\rho$ DER with center at $\rho$. On the other hand, the partial deviations, mentioned in 1(b), cannot be generalized to outline a total $\rho$ DER.

To study the influence of both inaccuracies and uncertainties on $\rho$ differential error, we considered a rearrangement of the terms in d$\rho$. The four inaccuracy terms corresponding to dm, $dm_k$ were defined as the di sum, and the three uncertainty terms corresponding to $d\rho_k$, as the du sum. After that, d$\rho$ was considered as a sum of two parts

$$d\rho = di + du \quad (14)$$

These conclusions can be applied to any other physical quantity, differentiably dependent on all, some or just one of the above independent variables. Thus, any such quantity has an L-term DER, where $7 \geq L \geq 1$. For example, the impedance Z of a DUT has a 7-term DER through

$$dZ = 2Z_0 d\rho/(1 - \rho)^2 = \zeta d\rho \quad (15)$$

$$dZ = \zeta(di + du) = dI + dU \quad (16)$$

that is a Z DER which results by stretching and rotating d$\rho$ with $\zeta = 2Z_0/(1 - \rho)^2$, so that, finally, the Z DER is similar to the $\rho$ DER.

## 3. Results

Although the developed expressions are independent of the particular measurement system in use, we report, for the sake of completeness, that measurements appearing in this paper were made using a type-N, $Z_0 = 50 \Omega$ measurement system with the following specific devices:

(i) HP8505A Opt 005PL VNA with (ii) Opt 007 HP8501A Storage Normalizer, (iii) HP8660C Synthesized Signal Generator with (iv) HP86603 RF Section, (v) HP5340A Opt 011 frequency counter, (vi) a HP85032A 50 Ω Type-N Calibration Kit, (vii) HP85032-60011 Open/Short, (viii) HP 8502 Transmission/Reflection Test Set, (ix) HP11501A -183 cm RF Cable, and (x) HP-IB IEEE488 82335B/8-bit ISA Interface Card under the control of an AMD486/66 PC.

This system operates from 1 to 1300 MHz with 100 Hz PLL stability in CW (non-sweep) frequency mode.

The set of standards used consists of a Short-circuit with $\rho_{k=1} = -1$, a matching Load with $\rho_{k=2} = 0$ and an Open-circuit with $\rho_{k=3} = +1$. These are the commonly SLO calibration standards given in Tab. 1.

| $\rho_k$ | $|\rho_k|$ | $\theta_k°$ |
|---|---|---|
| A | 1 | 180 |
| B | 0 | - |
| C | 1 | 0 |

**Tab.1.** SLO standard reflection coefficient values.

Substitution of the SLO reflection coefficient values to (5)-(11), simplifies the developed expressions for errors and differentials:

$$D = b \qquad (17)$$

$$M = (c + a - 2b)/(c - a) \qquad (18)$$

$$R = 2(a - b)(b - c)/(c - a) \qquad (19)$$

$$dD = -[2(a - b)(b - c)/(c - a)]dB + db \qquad (20)$$

$$dM = [(a - b)/(c - a)]dA$$
$$+ [4(b - c)(a - b)/(c - a)^2]dB$$
$$+ [(b - c)/(c - a)]dC$$
$$- [2(b - c)/(c - a)^2]da$$
$$- [2/(c - a)]db$$
$$- [2(a - b)/(c - a)^2]dc \qquad (21)$$

$$dR = [(a - b)(b - c)/(c - a)]dA$$
$$- [2(b - c)^2/(c - a)^2]da$$
$$+ \{4[(c - a) + 2(b - c)](a - b)(b - c)/(c - a)^2\}dB$$
$$- \{2[(c - a) + 2(b - c)]/(c - a)\}db$$
$$- [(a - b)(b - c)/(c - a)]dC$$
$$+ [2(a - b)^2/(c - a)^2]dc\} \qquad (22)$$

On the occasion, the specific $\rho_k$ mentioned in 1(b) are the SLO values that result (17)-(22), which obviously cannot be generalized to express total differential errors in any other case.

Therefore, as the matching load $\rho = 0$ is included in any SLO calibration, such a measurement system has a differential error core consists of a D DER with 4 line segments, 4 circular arcs and 8 vertices at most, an R DER and M DER with 20 line segments, 20 circular arcs and 40 vertices at most. Consequently, such a system produces a $\rho$ DER and a Z DER with 24 line segments, 24 circular arcs and 48 vertices at most.

The considered load uncertainties are given in Tab. 2, where in the absence of manufacturers' data for dA and dC, both of them were considered equal to the uncertainty of the Maury 8810B1 Open-circuit [7].

| $d\rho_k$ | $d|\rho_k|$ | $d\theta_k$ | |
|---|---|---|---|
| dA | 0 | 0.010 | -180 / -178 |
| | | | +178 / +180 |
| dB | 0 | 0.029 | - |
| dC | -0.010 | 0 | -2 / +2 |

**Tab. 2.** Intervals of SLO uncertainty values.

The annular region for any VNA measurement is specified from −100 to 0 db in modulus and ±180° in argument. Measurements result with a decimal unnormalized floating-point mantissa of 4 digits, for both modulus and argument.

It is well known that VNA measurements are referenced to a test connector (reference plane), which can be either the test port itself on the Reflection/Transmission Test Set [3(viii)] or the far-end connector of a $Z_0$ Transmission Line [3(ix)], connected to this test port. In either case, the error model of the system is still that of Fig. 1, with different error values of course. These two possibilities were considered here as two system configurations: System 1 and System 2.

A suite of developed software applications:

(a) Controls the system and collects data in terms of frequency, using the IEEE-488 protocol,

(b) Processes the collected data and computes the vertices of DER and the end-points of its DEIs and

(c) Sketches pictures for D, M, R, $\rho$, Z and their DERs in terms of frequency steps and makes a film using them as frames.

### 3.1 System Errors

The measurements of two System configurations were processed in different ways to demonstrate the variety of possible measurement considerations.

System 1: Measurements were made from 2 to 1289 MHz in 13 MHz steps. Each load $\rho_k$ was measured twice and the mean value of these measurements was considered as $m_k$. The endpoints of inaccuracy intervals $dm_k$ were considered as the two signed values of the absolute half difference between the two measurements, plus 1/2 of the unit in the last place of the mantissa, both in modulus and argument. The centre frequency $f_1 = 639$ MHz of the band was selected to reveal DER details. The resulting $m_k$ and $dm_k$ are given in Tab. 3.

| 1 | $|m_k|$ db | $\varphi_k°$ | 1 | $d|m_k|$ db | $d\varphi_k°$ |
|---|---|---|---|---|---|
| a | -0.625 | -178.8 | da | 0.020 | 2.075 |
| b | -49.8 | 3.95 | db | 0.050 | 7.300 |
| c | -0.5 | 2 | dc | 0.025 | 1.650 |

**Tab. 3.** System 1: The considered measurements and inaccuracies at $f_1$.

System 2: Measurements were made from 600 to 1000 MHz in 4 MHz steps. Each load $\rho_k$ was measured once. Although, a detail scale of inaccuracy in terms of signal level is available, the measurement inaccuracies $dm_k$ were considered here as symmetric intervals defined by 1 unit in the last place of the corresponding mantissa, both in modulus and argument of $m_k$. This emulates any other case in which there is no further information, so inaccuracy must be considered as the least inaccuracy, independent of the four possible rounding methods (to nearest, down, up or towards zero). In other words, this measurement consideration results the differential error core of the System 2. The frequency $f_2 = 932$ MHz was selected to detail the proposed method. The measurements and the considered inaccuracies at $f_2$ are given in Tab. 4.

| 2 | $|m_k|$ db | $\varphi_k°$ | 2 | $d|m_k|$ db | $d\varphi_k°$ |
|---|---|---|---|---|---|
| a | -1.47 | 122 | da | 0.01 | 1 |
| b | -25.0 | 44.9 | db | 0.1 | 0.1 |
| c | -1.40 | -43.5 | dc | 0.01 | 0.1 |

**Tab. 4.** System 2: The considered measurements and inaccuracies at $f_2$.

Tab. 5 contains the comparison of the errors in the two Systems at the selected frequencies. The wide diversity between the errors of System 1 and System 2 against the frequency results immediately from the comparison of Fig. 2 with Fig. 4. Fig. 3 shows the error DERs for the two Systems 1 and 2 at $f_1$ and $f_2$ respectively, with different scaling for the two coordinate axes to reveal the details. The contours are outlined with small circles as their vertices.

Since the objective was the uncertainty of antenna impedance and the related application was studied using System 2, additional example calculations are done at $f_2$ for an appropriately selected contour point, which is marked with an arrow on the related DERs below. The rectangular form of data for this example is given in Tab. 6.

| | $|D|$ db | $\varphi_D°$ | $|M|$ db | $\varphi_M°$ | $|R|$ db | $\varphi_R°$ |
|---|---|---|---|---|---|---|
| 1 | -49.8 | 3.95 | -42.16 | 61.28 | -0.562 | 1.60 |
| 2 | -25.0 | 44.9 | -24.21 | 80.0 | -1.474 | -50.8 |

**Tab. 5.** System errors at the selected frequencies.

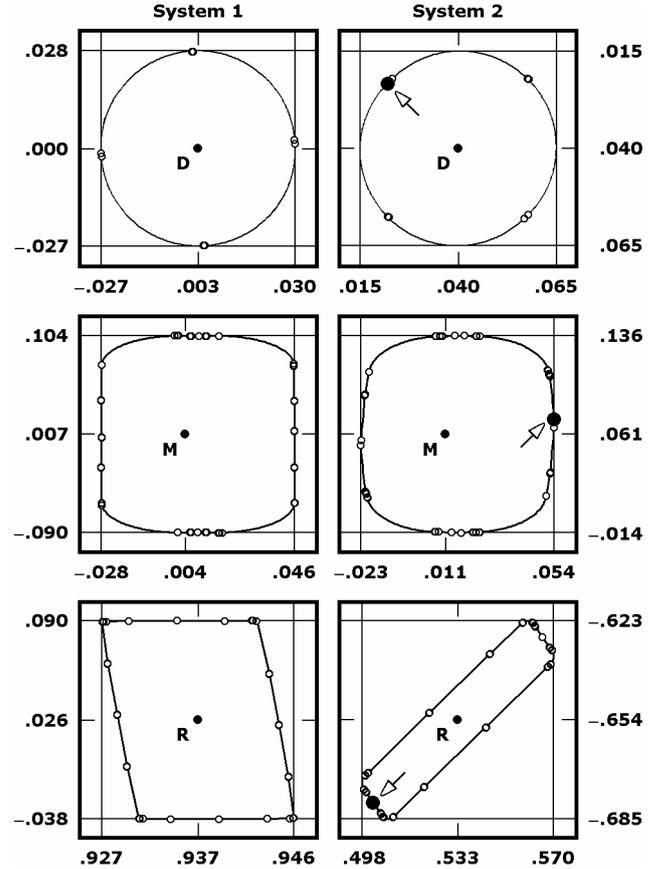

**Fig. 3.** System error DERs at the selected frequencies.

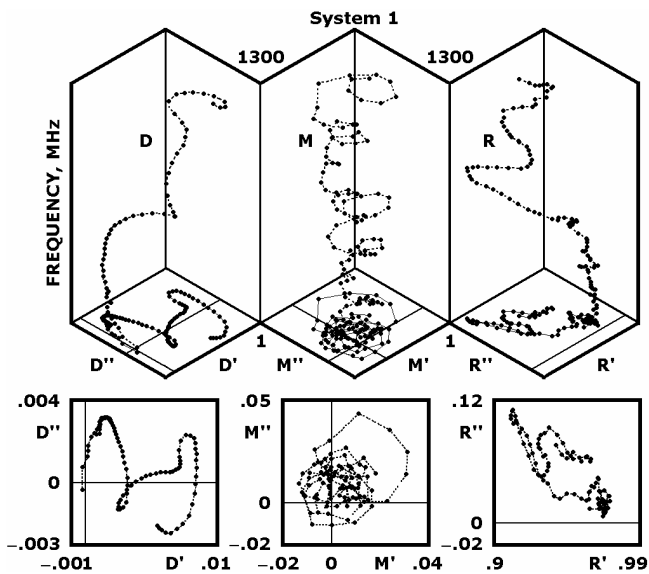

**Fig. 2.** System 1 errors against frequency.

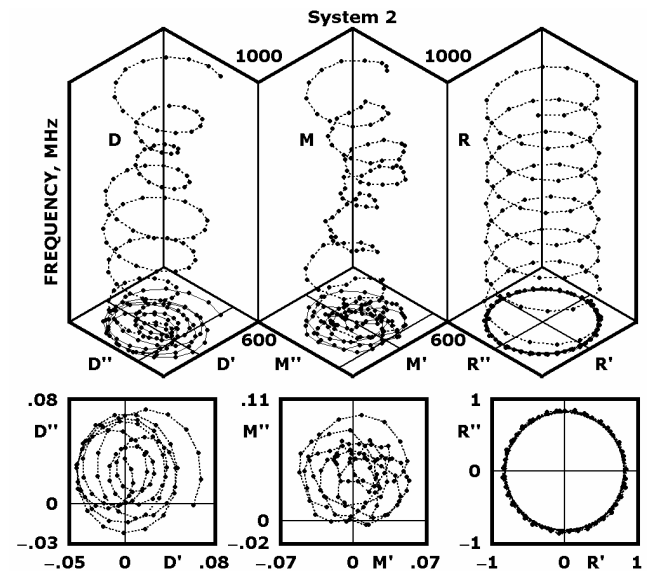

**Fig. 4.** System 2 errors against frequency.

| 2 | Re | Im | 2 | Re | Im |
|---|---|---|---|---|---|
| A | -1 | 0 | dA | -0.0100 | -0.0349 |
| B | 0 | - | dB | 0.0289 | 0.0029 |
| C | 1 | 0 | dC | 0.0100 | 0.0349 |
| a | -0.4474 | 0.7160 | da | 0.0130 | 0.0070 |
| b | 0.0398 | 0.0397 | db | -0.0005 | -0.0004 |
| c | 0.6174 | -0.5859 | dc | -0.0003 | -0.0018 |
| D | 0.0398 | 0.0397 | dD | -0.0178 | 0.0169 |
| M | 0.0106 | 0.0607 | dM | 0.0429 | 0.0112 |
| R | 0.5335 | -0.6540 | dR | -0.0317 | -0.0256 |

**Tab. 6.** System 2: Example results for a contour point.

## 3.2 An Antenna in contrast with a Resistor

DUT 1: A typical resistor with a nominal DC impedance of 50 Ω ±20% tolerance was soldered on a type-N base connector and enclosed in an aluminum box to form an EM shielded DUT for reference.

DUT 2: A typical UHF ground-plane antenna of five λ/4 elements at 900 MHz, with apex angle 90°, was built by copper bare wire of 1 mm diameter and its terminals were soldered directly on a type-N connector of a rather poor dielectric insulation. Therefore, it is, in essence, an EM unshielded DUT. The antenna was roughly installed indoors, nearby and outside of an anechoic chamber.

The antenna was simulated by 96 wire segments. The simulation was carried out with a suite of developed visual tools supported by a fully analyzed, corrected and redeveloped edition of the original thin-wire computer program by Richmond [8] while the connector was simulated separately.

The measurements m and their inaccuracies dm of 50 Ω DC Resistor at $f_1$ and UHF Ground-Plane Antenna at $f_2$, were considered under the mentioned measurement conditions for System 1 and System 2 and they are given in Tab. 7 and Tab. 8, respectively.

| 1 | |m| db | φ° | 1 | d|m| db | dφ° |
|---|---|---|---|---|---|
| m | -10.4 | -21.75 | dm | 0.050 | 0.200 |

**Tab. 7.** 50 Ω DC Resistor: Measurement and inaccuracy at $f_1$.

| 2 | |m| db | φ° | 2 | d|m| db | dφ° |
|---|---|---|---|---|---|
| m | -8.21 | -155 | dm | 0.01 | 1 |

**Tab. 8.** UHF Ground-Plane Antenna: Measurement and inaccuracy at $f_2$.

The values in Tab. 9 complete the example of the selected contour point for UHF Ground-Plane Antenna at $f_2$.

The precise relation of the total complex differential error dρ to its complex differential error parts di and du, due to all inaccuracies and all uncertainties respectively, is illustrated by their DERs in Fig. 5 for the 50 Ω DC Resistor at $f_1$ and in Fig. 6 for the UHF Ground-Plane Antenna at $f_2$.

| 2 | Re | Im | 2 | Re | Im |
|---|---|---|---|---|---|
| m | -0.3522 | -0.1642 | dm | -0.0033 | -0.0060 |
| ρ | -0.0975 | -0.4989 | dρ | 0.0694 | -0.0030 |
| Z | 25.5 | -34.3 | dZ | 3.0 | -3.7 |

**Tab. 9.** UHF Ground-Plane Antenna: Results for the contour point.

The circumscribed dash-dotted circle to each DER, corresponds to the max value of |dρ|, |di| and |du|, respectively.

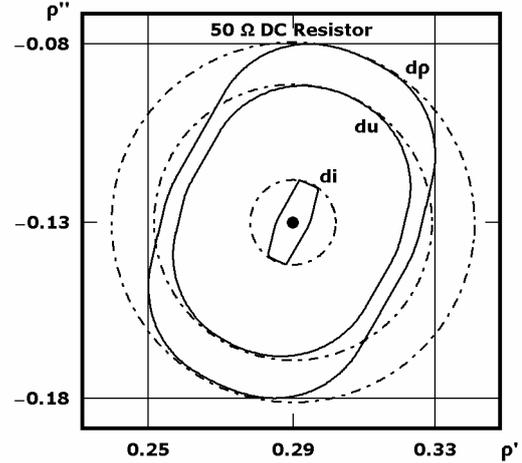

**Fig. 5.** 50 Ω DC Resistor: ρ related DERs at $f_1$.

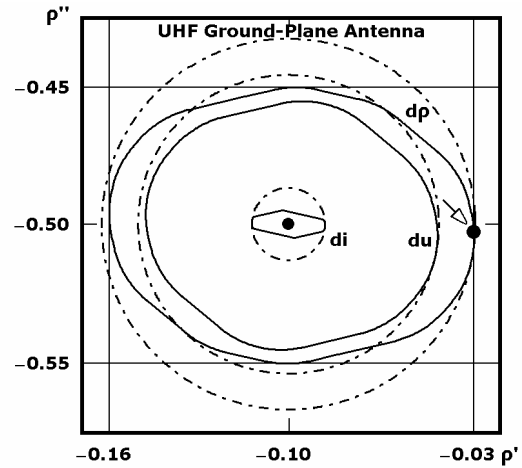

**Fig. 6.** UHF Ground-Plane Antenna: ρ related DERs at $f_2$.

The illustrations for the uncertainty dZ = dI + dU and the calculated exact difference ΔZ for 50 Ω DC Resistor at $f_1$ and UHF Ground-Plane Antenna at $f_2$, are shown in Fig. 7 and Fig. 10, respectively. Numeric evaluation of ΔZ was resulted $2^{7 \times 2}$ points, from L = 7 interval endpoints for dm, $dm_k$ and $dρ_k$, which are dense enough to appear as stripes, placed over the Z DER. The computation time for ΔZ calculations exceeds that for Z DER by a factor of about 60. It is concluded that almost all ΔZ points belong to Z DER.

The precise relation of the Z DER to its complex differential error parts dI and dU, geometrically represented by their DERs, make clear that measurement inaccuracies are not insignificant in Z uncertainty calculations.

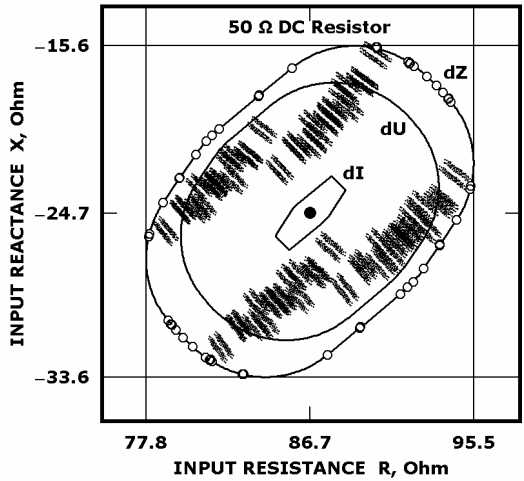

**Fig. 7.** 50 Ω DC Resistor: Z related DERs and ΔZ at $f_1$.

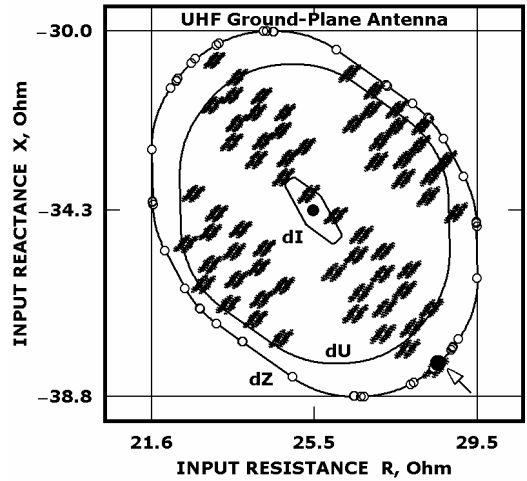

**Fig. 10.** UHF Ground-Plane Antenna: Z related DERs and ΔZ at $f_2$.

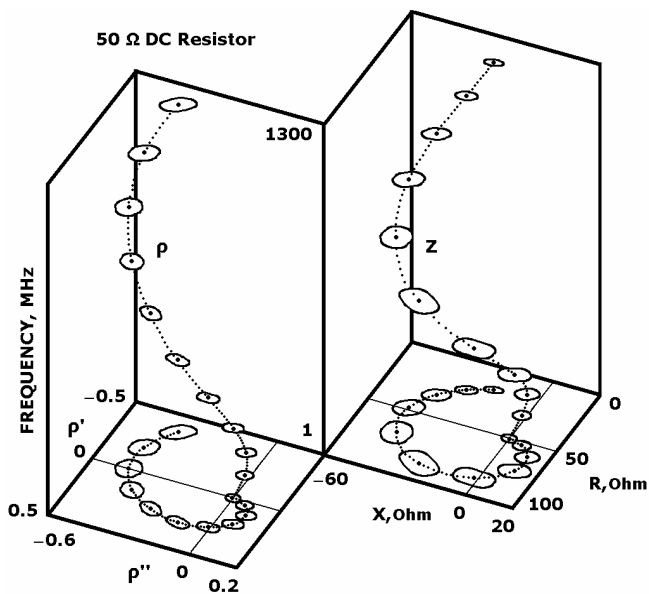

**Fig. 8.** 50 Ω DC Resistor: ρ and Z DERs against frequency.

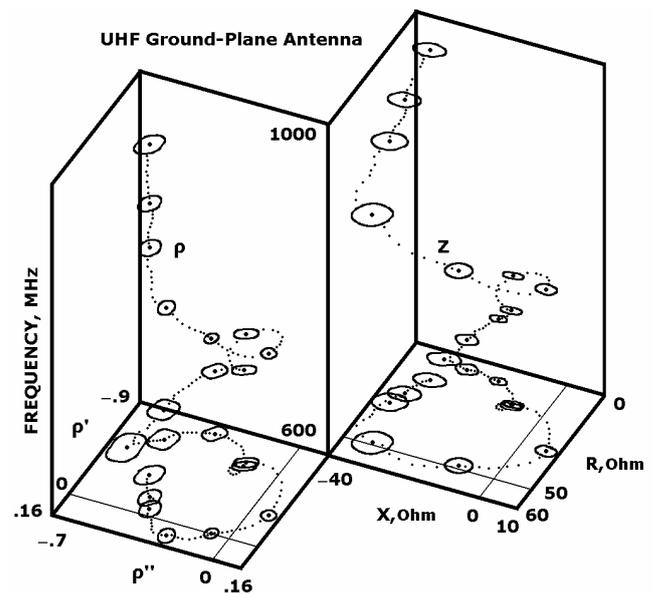

**Fig. 11.** UHF Ground-Plane Antenna: ρ and Z DERs against frequency.

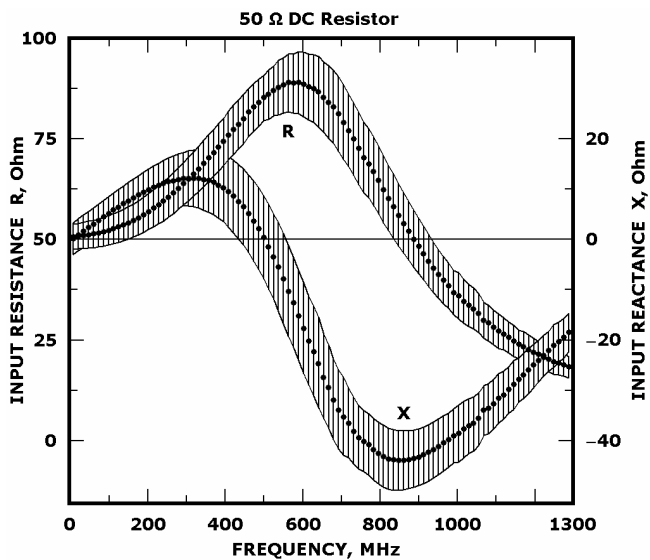

**Fig. 9.** 50 Ω DC Resistor: Z-DEIs against frequency.

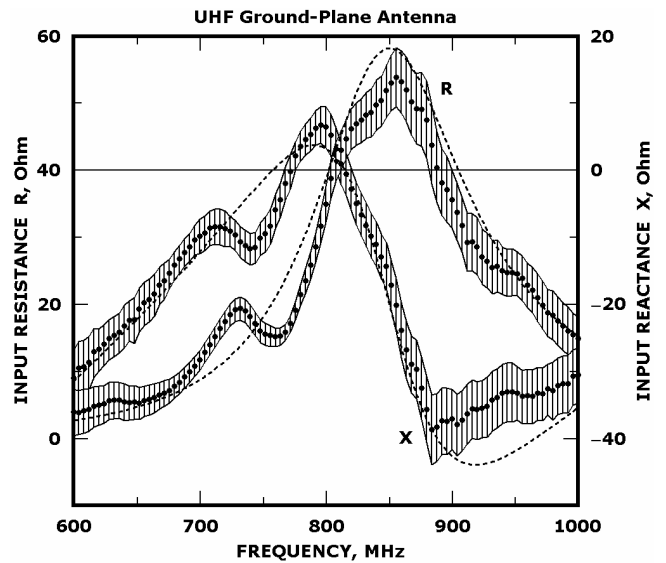

**Fig. 12.** UHF Ground-Plane Antenna: Z-DEIs against frequency.

To demonstrate the method, selected DER frames for ρ and Z, mentioned at 3(c), are shown, as beads on space-curved filaments against frequency, in Fig. 8 for 50 Ω DC Resistor and Fig. 11 for UHF Ground-Plane Antenna.

The computed DEIs for the input resistance R and reactance X against frequency are shown, in Fig. 9 for 50 Ω DC Resistor and Fig. 12 for UHF Ground-Plane Antenna.

In Fig. 12, the dashed lines represent predicted values, for the input impedance Z of UHF Ground-Plane Antenna simulation, which are closed enough to the computed DEIs.

In Fig. 13, the ρ DER is compared with the approximate $|\Delta S_{11}|$. It is concluded that $\Delta S_{11}$ underestimates the uncertainty of 50 Ω DC Resistor at $f_1$ and overestimates that of UHF Ground-Plane Antenna at $f_2$.

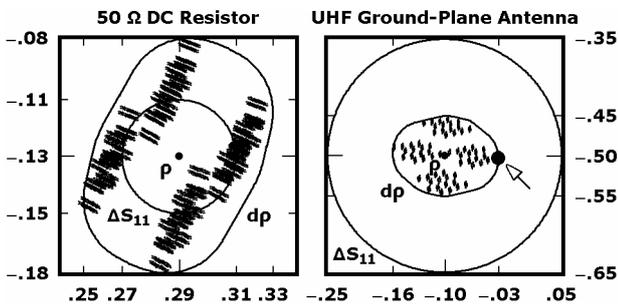

**Fig. 13.** Precise comparison between Δρ stripes, ρ-DER and $\Delta S_{11}$ circle.

To estimate roughly the separate contribution of all inaccuracies and all uncertainties to the differential error of ρ and Z, max values are commonly used. In Tab. 10 the max values of |dρ|, |di| and |du| are expressed as percentage of |dρ| from Fig. 5 and Fig. 6, and of $|\Delta S_{11}|$ from Fig. 13. Since from (15) and (16) |dZ|, |dI| and |dU| are analogous to |dρ|, |di| and |du| respectively, the max values of them are also given in the same columns of Tab. 10. Although max|dZ| ≤ max|dI| + max|dU|, the particular shape of the ρ related DERs, in Fig. 5 and Fig. 6, and of their similar Z related DERs, in Fig. 7 and Fig. 10, result in max|dZ| ≅ max|dI| + max|dU|, as shown in Tab. 10.

| Max | 50 Ω DC Resistor | | | UHF Ground-Plane Antenna | | | |
|---|---|---|---|---|---|---|---|
| | $|\Delta S_{11}|$ | Max | |dρ|/|dZ| | Max | $|\Delta S_{11}|$ | Max | |dρ|/|dZ| |
| |dρ| | 210 | |dZ| | | |dρ| | 45 | |dZ| | |
| |di| | 50 | |dI| | 25 | |di| | 10 | |dI| | 20 |
| |du| | 160 | |dU| | 75 | |du| | 40 | |dU| | 80 |

**Tab. 10.** Percentage comparison of max differential errors at $f_1$ and $f_2$.

Tab. 11 contains the results for max values of differential errors over the whole measurement bands of System 1 and System 2. Great divergences of $|\Delta S_{11}|$ from dρ are noted. Max|dI| contributes an amount of about 30 ~ 35% of max|dZ| to the total max|dZ|, under these rather conservative considerations for the particular applications. This result complements the one from Z related DERs of Fig. 7 and Fig. 10, so it is concluded that measurement inaccuracies are significant indeed in Z uncertainty calculations.

| Max | 50 Ω DC Resistor | | | UHF Ground-Plane Antenna | | | |
|---|---|---|---|---|---|---|---|
| | $|\Delta S_{11}|$ | Max | |dρ|/|dZ| | Max | $|\Delta S_{11}|$ | Max | |dρ|/|dZ| |
| |dρ| | 155~2660 | |dZ| | | |dρ| | 30~145 | |dZ| | |
| |di| | 35~195 | |dI| | 5~35 | |di| | 5~25 | |dI| | 5~30 |
| |du| | 120~2470 | |dU| | 65~95 | |du| | 25~135 | |dU| | 70~95 |

**Tab. 11.** Percentage comparison of max differential errors vs. frequency.

Hence, the proposed method may be efficiently used in any other case where the process equations (3), (4) and the defined DERs and DEIs can find application.

# References

[1] FITZPATRICK, J. *Error models for systems measurement.* Microwave Journal, May 1978, vol. 21, p. 63-66.

[2] SPIEGEL, M.R. *Complex Variables with an introduction to Conformal Mapping and its applications.* McGraw-Hill, 1974, p.203.

[3] - *Vector Measurements of High Frequency Networks.* Hewlett - Packard, 1989, p.3-6.

[4] BALLO, D. *Network Analyzer Basics.* Hewlett-Packard, 1998, p.1-61, 1-68.

[5] STUMPER, U. *Influence of TMSO calibration standards uncertainties on VNA S-parameters measurements.* IEEE Transactions on Instrumentation and Measurements, April 2003, vol. 52, no. 2, p. 311 - 315.

[6] CHIPMAN, R.A. *Transmission Lines.* McGraw-Hill, 1968, p.137.

[7] - *Precision Microwave Instruments and Components.* Maury Microwave Corporation, 1996, p.132.

[8] RICHMOND, J.H. *Radiation and scattering by thin-wire structures in a homogeneous conducting medium.* IEEE Transactions on Antennas and Propagation, Vol. 22, Issue 2, March 1974, p.365

# About Authors...

**Nikolitsa YANNOPOULOU** was born in Chania, Crete, Greece in 1969. She graduated in 1992 from Electrical Engineering at Democritus University of Thrace, Xanthi, Greece and since then she is with Antennas Research Group at Democritus University. She received the MSc degree with full marks in Microwaves at Democritus University in 2003. She is now finishing her PhD at the same University. Her research interests are in antenna theory, software, built, measurements and virtual laboratories.

**Petros ZIMOURTOPOULOS** was born in Thessaloniki, Greece in 1950. He received his MSc degree from Aristotle University of Thessaloniki, in 1978 and his PhD degree from Democritus University of Thrace in 1985. He is Assistant Professor in Electrical Engineering and Computer Engineering at Democritus University. He is the leader of Antennas Research Group, which he founded in 1985.